\documentclass[prl,twocolumn,showpacs,floatfix,superscriptaddress]{revtex4-1}
\usepackage[T1]{fontenc}
\usepackage{graphicx}
\usepackage{amstext}
\usepackage{amsxtra}
\usepackage{color}
\begin{document}
\title{Light storage in a room temperature atomic vapor based on coherent population oscillations}
%\shorttitle{Long-lived light storage based on coherent population oscillations} %Insert here a short version of the title if it exceeds 70 characters

\author{M.-A. Maynard}
\affiliation{Laboratoire Aim\'e Cotton, CNRS - Universit\'e Paris Sud 11 - ENS Cachan, 91405 Orsay Cedex, France}
\author{F. Bretenaker}
\affiliation{Laboratoire Aim\'e Cotton, CNRS - Universit\'e Paris Sud 11 - ENS Cachan, 91405 Orsay Cedex, France}
\author{F. Goldfarb}
\email{fabienne.goldfarb@u-psud.fr}
\affiliation{Laboratoire Aim\'e Cotton, CNRS - Universit\'e Paris Sud 11 - ENS Cachan, 91405 Orsay Cedex, France}


\date{\today}
\begin{abstract}
We report the experimental observation of Coherent  Population Oscillation (CPO) based light storage in an atomic vapor cell at room temperature.  Using the ultranarrow CPO between the ground levels of a $\Lambda$ system selected by polarization in metastable $^4$He, such a light storage is experimentally shown to be phase preserving. As it does not involve any atomic coherences it has the advantage of being robust to dephasing effects such as small magnetic field inhomogeneities. The storage time is limited by the population lifetime of the ground states of the $\Lambda$ system.
\end{abstract}

\maketitle

%\section{Section title}
Because they are essential for the development of many devices in quantum communication networks, optical memories have become a very active research topic in the area of quantum information processing. Different approaches have been developed to store light in atomic system excitations, such as photon-echo or Electromagnetically Induced Transparency (EIT) based memories \cite{Lvovsky2009}. In gas cells, high efficiencies were obtained in alkali atoms \cite{Novikova2012} -- mainly rubidium -- using EIT close to \cite{Phillips2001} or far-off optical resonance \cite{Reim2010}, Gradient Echo Memories (GEM) \cite{Hetet2008a} or four-wave mixing \cite{Camacho2009}. All these methods are based on the excitation of coherence between atomic levels. They can consequently be efficiently implemented only in systems in which these coherences have a long lifetime. The storage time and the efficiency are thus highly sensitive to all dephasing mechanisms such as, e.g., magnetic field inhomogeneities.

Another protocol based on long-lived Coherent Population Oscillation (CPO) was theoretically proposed to implement spatial optical memories \cite{Eilam2010}. CPO occurs in a two-level system (TLS) when two coherent electromagnetic fields of different amplitudes and frequencies drive the same transition. the beatnote between these fields leads to a temporal modulation of the excited and ground state population difference, with a bandwidth linked to the upper level population lifetime \cite{Schwartz1967,Boyd1981}. This lifetime, and thus the associated memory lifetime, can be increased by using a TLS whose upper level population decays via a shelving state \cite{Eilam2010}. In the present letter, we report what is to our knowledge the first experimental demonstration of storage based on CPO. Instead of a TLS assisted by shelving state, we use a $\Lambda$ system composed of two coupled TLSs: this gives rise to an ultranarrow CPO resonance due to the transfer of population modulations to CPOs between the lower states of the $\Lambda$ system \cite{Laupretre2012}. As it does not involve atomic coherences, it has the advantage to be robust to dephasing effects illustrated by small magnetic field inhomogeneities.

\begin{figure}[t]
\center
\includegraphics[width=0.46\textwidth]{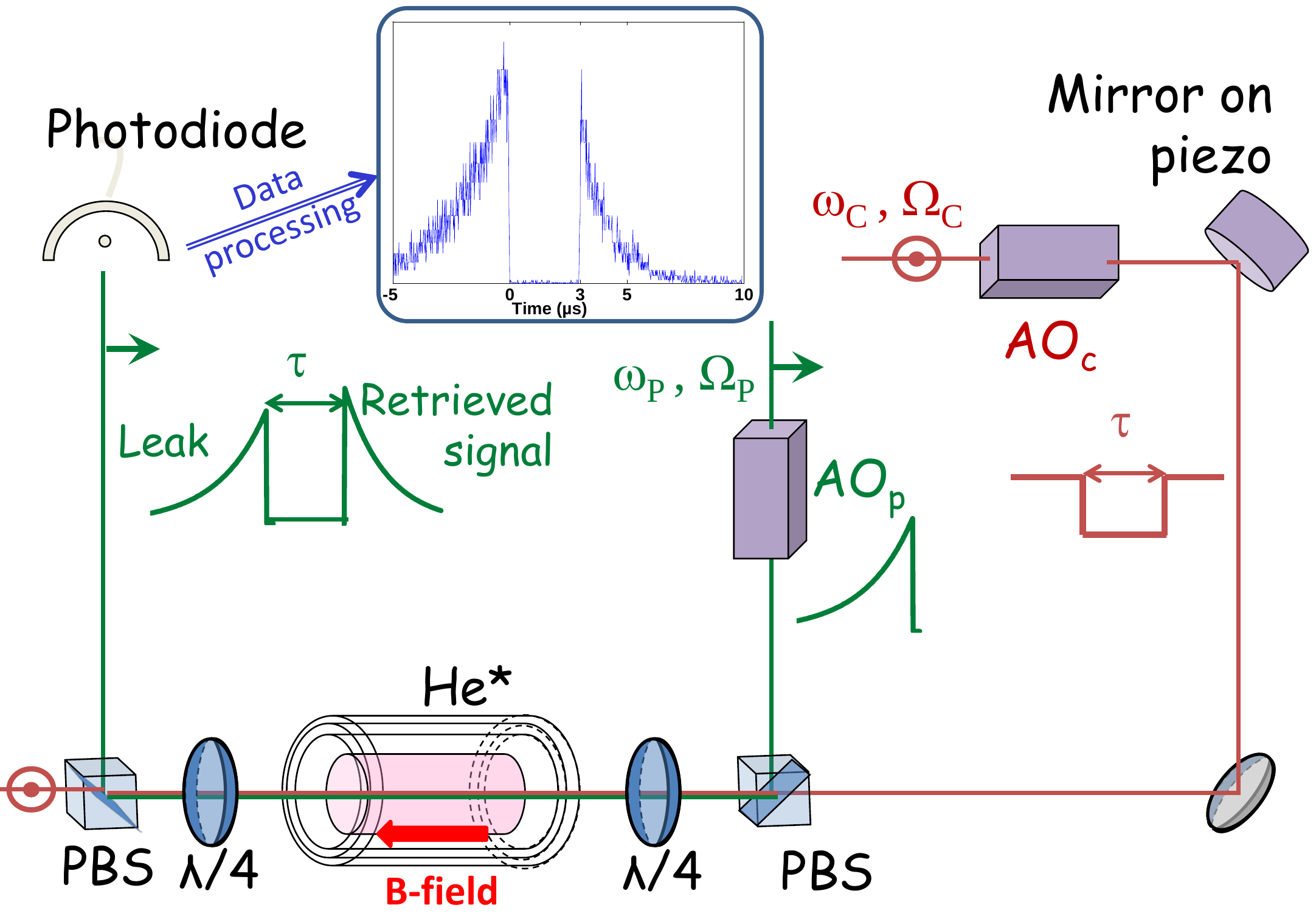} 
\caption{\label{fig1} Experimental setup for EIT or ultranarrow CPO storage in metastable $^4$He. The orthogonally polarized coupling and probe beams of optical frequencies $\omega_c$ and $\omega_p$ and Rabi frequencies $\omega_c$ and $\Omega_c$ and $\Omega_p$ respectively, are separated or recombined with polarizing beam splitters (PBS). They are controlled in frequency and amplitude by acousto-optic modulators (AO$_c$ and AO$_p$).  $\lambda/4$ plates can be added to generate circular polarizations (EIT configuration). A $\mu$-metal shielding protects the cell from stray magnetic fields. A solenoid can provide a longitudinal B-field. A piezoelectric transducer is used for homodyne detection. Inset: Recorded leak and retrieved signals after data processing.}
\end{figure}

\begin{figure}[t]
\includegraphics[width=0.47\textwidth]{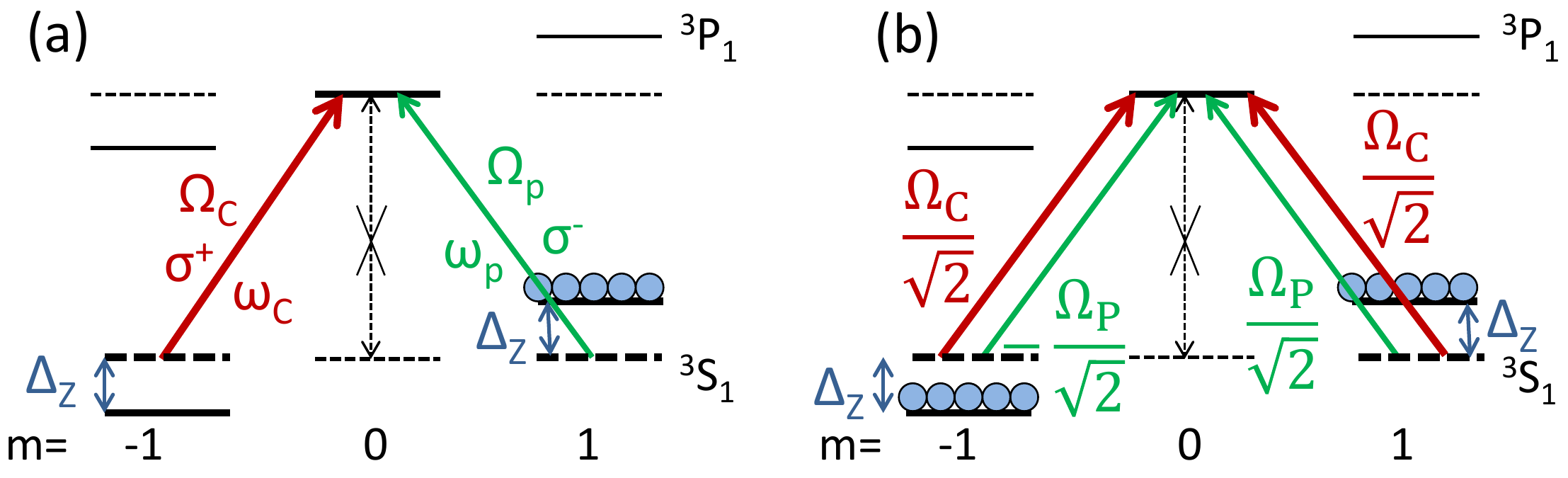}
\includegraphics[width=0.47\textwidth]{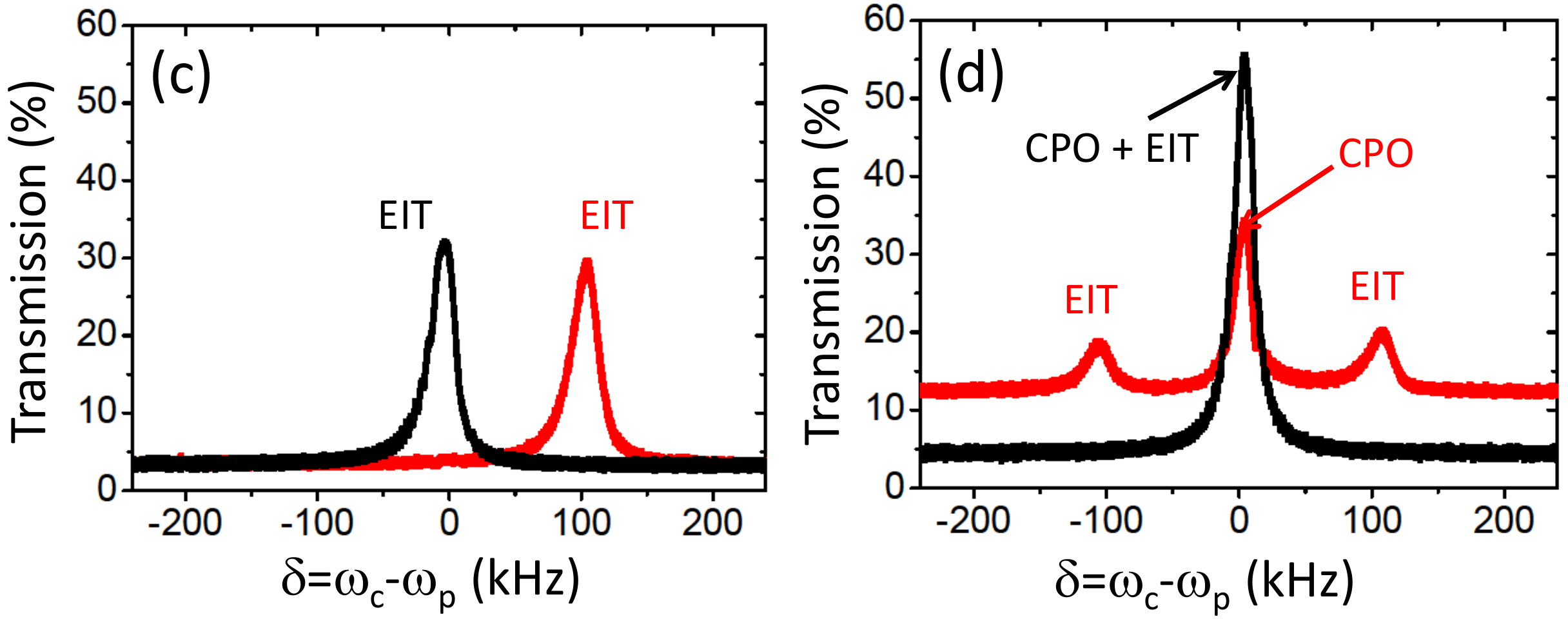}
\caption{\label{fig2} (a) circ\,$\perp$\,circ: $\sigma^+$ coupling beam of optical and Rabi frequencies $\omega_c$ and $\Omega_c$ and $\sigma^-$ probe beam of optical and Rabi frequencies $\omega_p$ and $\Omega_p$); (b) lin\,$\perp$ \,lin configurations with a magnetic field that lifts the Zeeman degeneracy by a quantity $\Delta_Z$. (c) circ\,$\perp$\,circ configuration: for a zero magnetic field, EIT occurs for coupling and probe beams of the same frequency (in black); For a 17 mG longitudinal magnetic field, the EIT window is shifted by $2\Delta_Z=100$\,kHz (in red). (d) lin\,$\perp$ \,lin configuration: at zero magnetic field,the resonance is due both to EIT and CPO (in black); With a longitudinal magnetic field, EIT resonances occur for $\pm 2\Delta_Z$ coupling and probe frequency detunings while the central transmission resonance (same coupling and probe frequency), where CPO-storage is performed, is due to ultranarrow CPO (in red) \cite{Laupretre2012}.}
\end{figure}

The experiment is based on the $2^3\mathrm{S}_1 \rightarrow 2^3\mathrm{P}_1$ (D1) transition of helium, that permits one to isolate a pure $\Lambda$ system involving only electronic spins \cite{Laupretre2012,Goldfarb2009}. The experimental setup is shown in  Fig.\,\ref{fig1}. The helium cell is 6 cm long and has a diameter of 2.5 cm. It is filled with 1 Torr of $^4$He and placed into a three-layer $\mu$-metal shield to remove magnetic fields gradients. It can be translated inside the shielding to induce more or less such inhomogeneities. The Doppler broadened transition half-width at half-maximum is about 0.9 GHz, but the optical pumping is effective over approximately half of the Doppler profile. Helium atoms are excited to the metastable state by a RF discharge at 27 MHz. Depending on the RF discharge, the linear transmission of a small probe is measured to lie between 0.1\% and 0.15\%. The 3\,mm diameter probe and coupling beams are derived from the same laser at 1083\,nm. They are controlled in frequency and amplitude by two acousto-optic modulators. The power of the coupling beam is set between 14 and 17 mW, which is equivalent to a coupling Rabi frequency $\Omega_C/2\pi$ between 28 and 30\,MHz. The probe beam power is about 90 $\mu$W. In these conditions, the delays associated to the CPO or EIT transmission resonances are 1 to 2 $\mu$s long.\\
An adjustable longitudinal magnetic field $B$generated by a solenoid lifts the degeneracy between the Zeeman sublevels by a quantity $\Delta_Z=g \mu_B B$, so that the ground levels are separated by $2\Delta_Z$ [see Fig.\,\ref{fig2}(a)]. $\mu_B$ is the Bohr magneton and as the Land\'e Factor for levels $2^3\mathrm{S}_1$ and $2^3\mathrm{P}_1$ is $g=2$, we have $2\Delta_Z=5.6$\,MHz/Gauss.

In the usual configuration for EIT along the D1 transition, the pump and probe beams are circularly and orthogonally polarized (circ\,$\perp$\,circ configuration) \cite{Goldfarb2009}. Since the $m=0 \rightarrow m=0$ transition is forbidden, a $\sigma^+$ coupling beam pumps the atoms into the $m=+1$ ground-state sublevel, which is probed by a $\sigma^-$  beam [see Fig.\,\ref{fig2}(a)]. As EIT occurs at Raman resonance, for equal coupling and probe optical detunings, a longitudinal B-field shifts the two-photon resonance by a frequency $2\Delta_Z$ [see Fig.\,\ref{fig2}(c)]. 

When the coupling beam is linearly polarized (lin\,$\perp$\,lin configuration), it excites both transitions of the $\Lambda$ system and atoms are equally pumped into both $m=\pm1$ sublevels [see Fig.\,\ref{fig2}(b)]. A perpendicularly polarized probe beam which couples both arms then exhibits two EIT resonances that can be shifted by $\pm 2\Delta_Z$ with a longitudinal magnetic field. Moreover, in this case, we showed previously that an ultra-narrow CPO resonance appears for equal coupling and probe frequencies \cite{Laupretre2012} [see Fig.\,\ref{fig2}(d)].

\begin{figure}[t]
\includegraphics[width=0.23\textwidth]{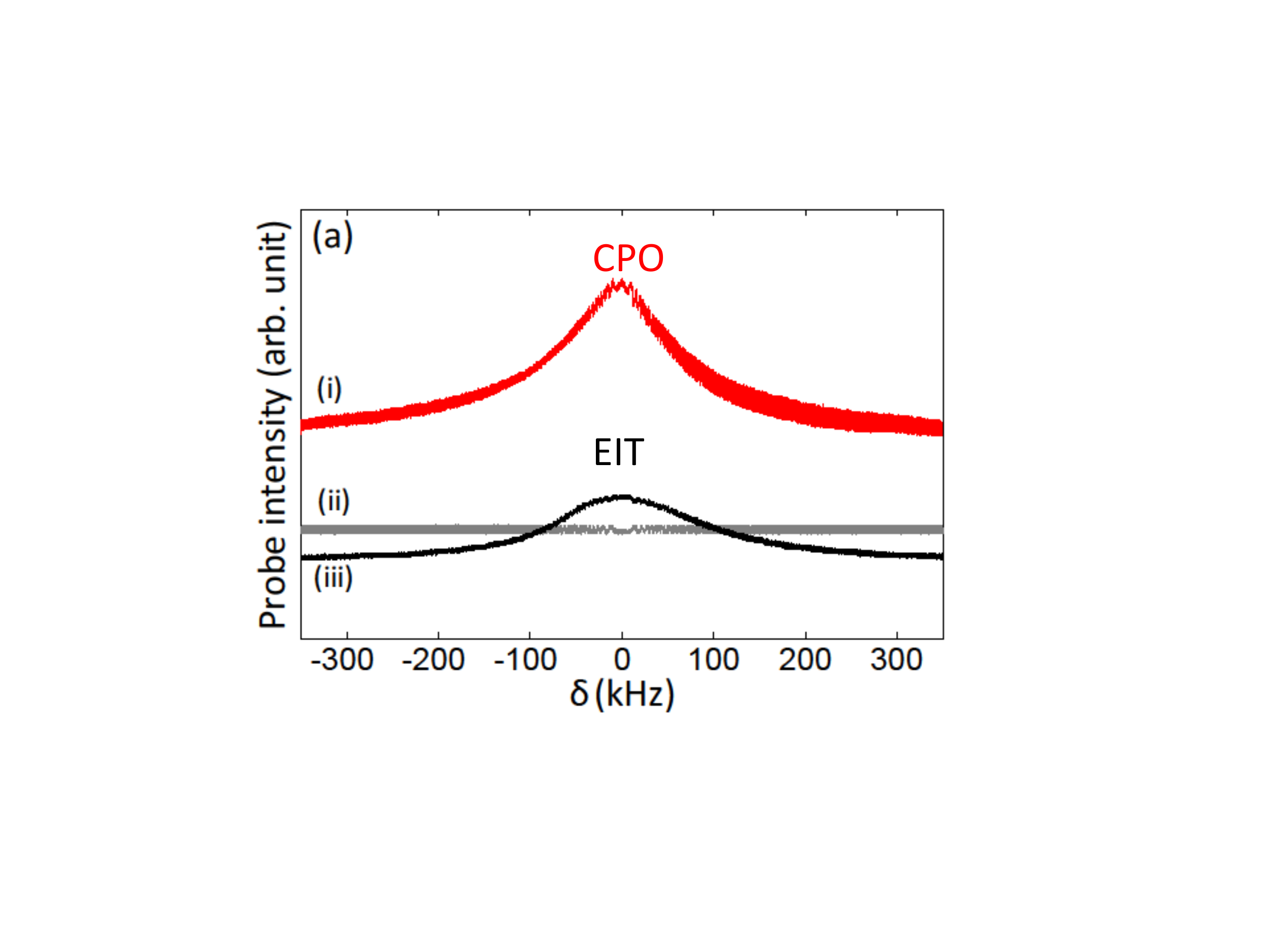}
\includegraphics[width=0.23\textwidth]{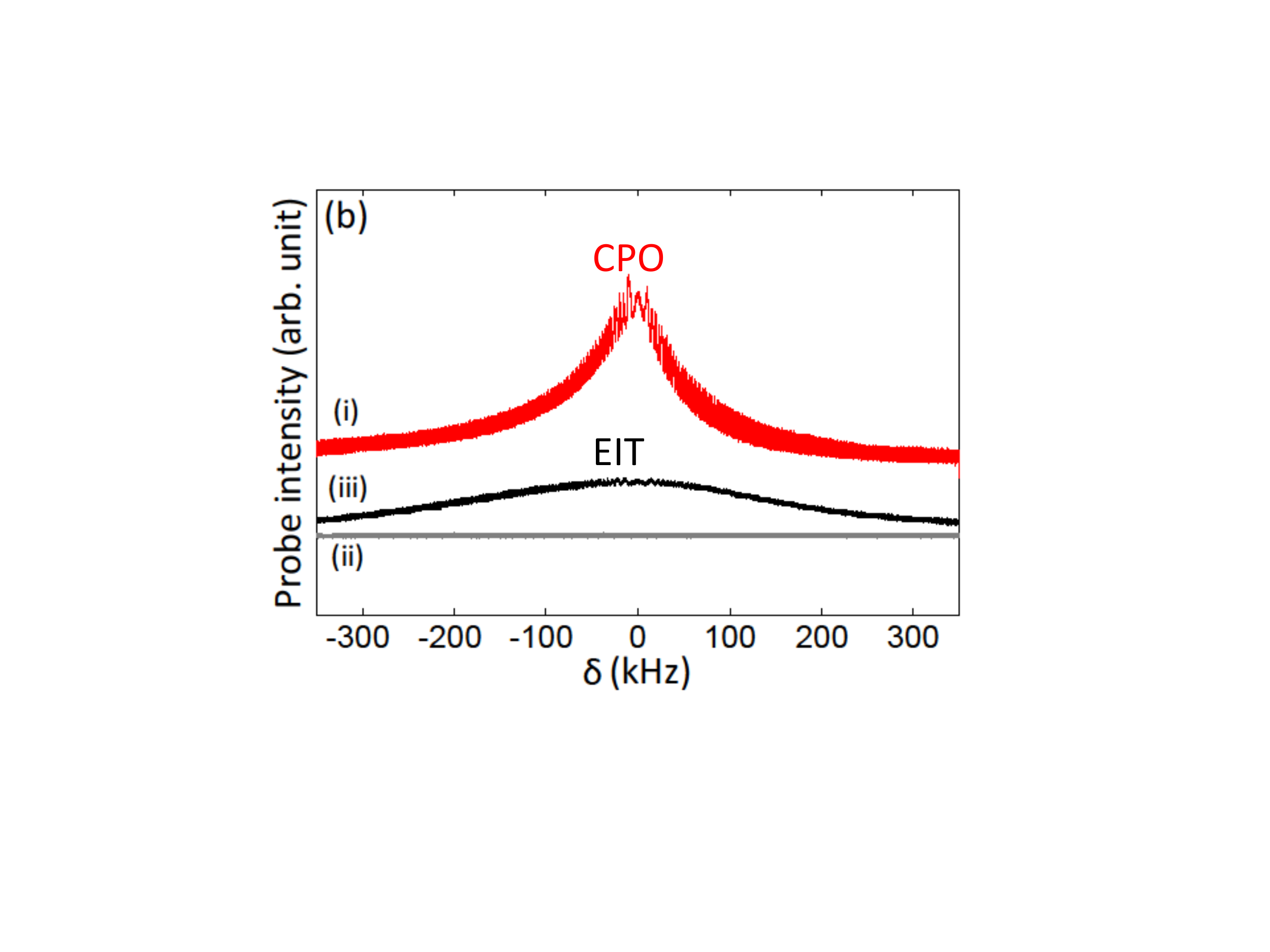}
\caption{\label{fig3} Transmission profiles (a) inside the $\mu$-metal and (b) at the edge of the $\mu$-metal shielding in: (i) the lin\,$\perp$ \,lin configuration and a 0.7\,Gauss longitudinal magnetic field  (in red); (ii) the circ\,$\perp$\,circ configuration  and a 0.7\,Gauss longitudinal magnetic field (in grey); (iii) the circ\,$\perp$\,circ configuration and no longitudinal magnetic field (in black). In case (ii), there is no EIT resonance for coupling and probe beams of the same frequency: in case (i) with linear polarizations, the resonance is thus only due to CPO.}
\end{figure}

The linewidth of this CPO resonance is much narrower than usual CPO resonances, which are limited by the population decay rate of the upper level $\Gamma_0$. Theoretical models derived in reference \cite{Laupretre2012} show that in this case, CPOs occur between the ground levels: their bandwidth is then limited by the population decay rate of these levels, e.g., the transit time of the atoms through the laser beam, instead of the population decay rate of the upper level. The width of such resonances is unaffected by a decrease of the coherence lifetime (induced for example by magnetic field gradients) that would on the contrary enlarge EIT resonances. Figs.\,\ref{fig3}(a) and \ref{fig3}(b) show transmission profiles recorded in both circ\,$\perp$\,circ and lin\,$\perp$\,lin configurations, respectively at the center and the edge of the $\mu$-metal shielding. $\delta$ is the detuning between the probe and coupling beams. The upper and red profile (i) is a CPO resonance obtained with linear polarizations and a longitudinal magnetic field of about 0.7\,Gauss: the EIT resonances are shifted by nearly 4\,MHz and are not visible in the probed window. The lower grey transmission profile (ii) is obtained with circular polarizations in the same conditions: as the EIT resonance is equally shifted, the absorption is flat. The black resonance (iii) is an EIT one obtained with circular polarizations and no added B-field (more precisely, a very small compensation longitudinal B-field of about 0.01\,Gauss is added when the cell is at the side of the $\mu$-metal shielding, so that the ground level remains degenerate and EIT occurs for the same coupling and probe frequencies). Its width is clearly increased by magnetic field inhomogeneities when the cell is at the edge of the shielding [see the black line in Fig.\,\ref{fig3}(b)], while the CPO resonance remains insensitive to a decrease of coherence lifetimes. In this position, the magnetic field is very inhomogeneous, but its typical magnitude is only a few tens of milligauss.

\begin{figure}[t]
\center
\includegraphics[width=0.32\textwidth]{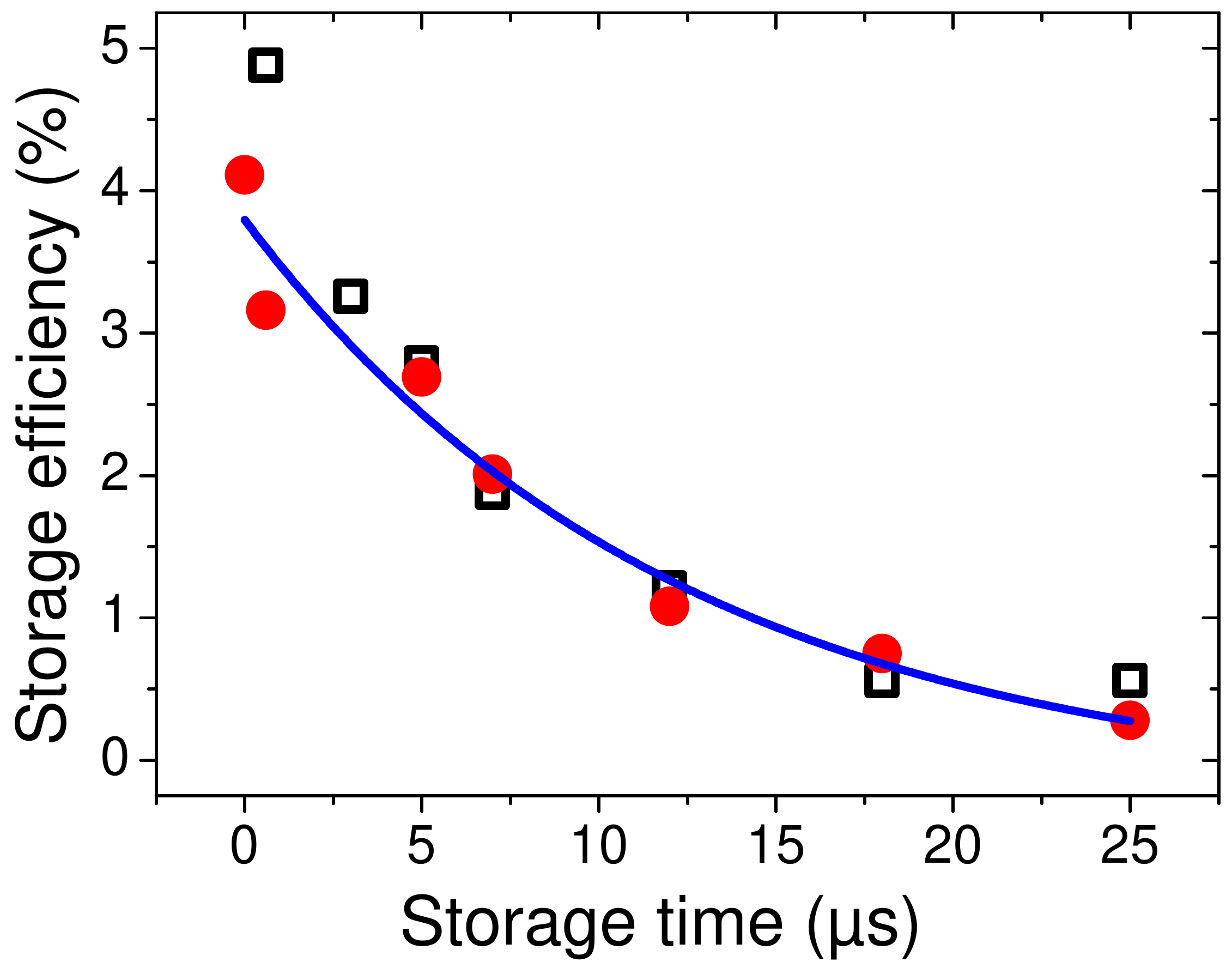}  
\caption{\label{fig4}Measured storage efficiency as a function of the storage time, for a 4\,$\mu$s rise-time exponential pulse, when the cell is inside the $\mu$-metal shielding. CPO storage efficiencies (red dots) are recorded in the  lin\,$\perp$\,lin configuration with a B=0.7\,Gauss longitudinal magnetic field. EIT storage (open black squares) measurements are performed in the circ\,$\perp$\,circ configuration. The full line is an exponential fit with a 10\,$\mu$s decay time constant.}
\end{figure}

The CPO storage experiments are performed in the lin\,$\perp$\,lin configuration [see Fig.\,\ref{fig2}(b)], using the central transmission window which appears when a magnetic field is added [see the red curve in Fig.\,\ref{fig2}(d)]. They are compared to EIT storage experiments performed in the circ\,$\perp$\,circ configuration [see Fig.\,\ref{fig2}(a)]. The 0.7\,Gauss magnetic field allows us to completely remove EIT storage by shifting the EIT resonances $\pm 4$\,MHz away. We use the same storage sequence for CPO and EIT storage. After switching on the coupling beam, the probe beam is progressively turned on with an exponential shape, followed by an abrupt decrease. Once the pulse has entered the helium cell, the coupling beam is suddenly switched off. After a storage time $\tau$ that can be varied, the coupling beam is switched on again and the retrieved pulse is released.

\begin{figure}[t]
\center 
\includegraphics[width=0.32\textwidth]{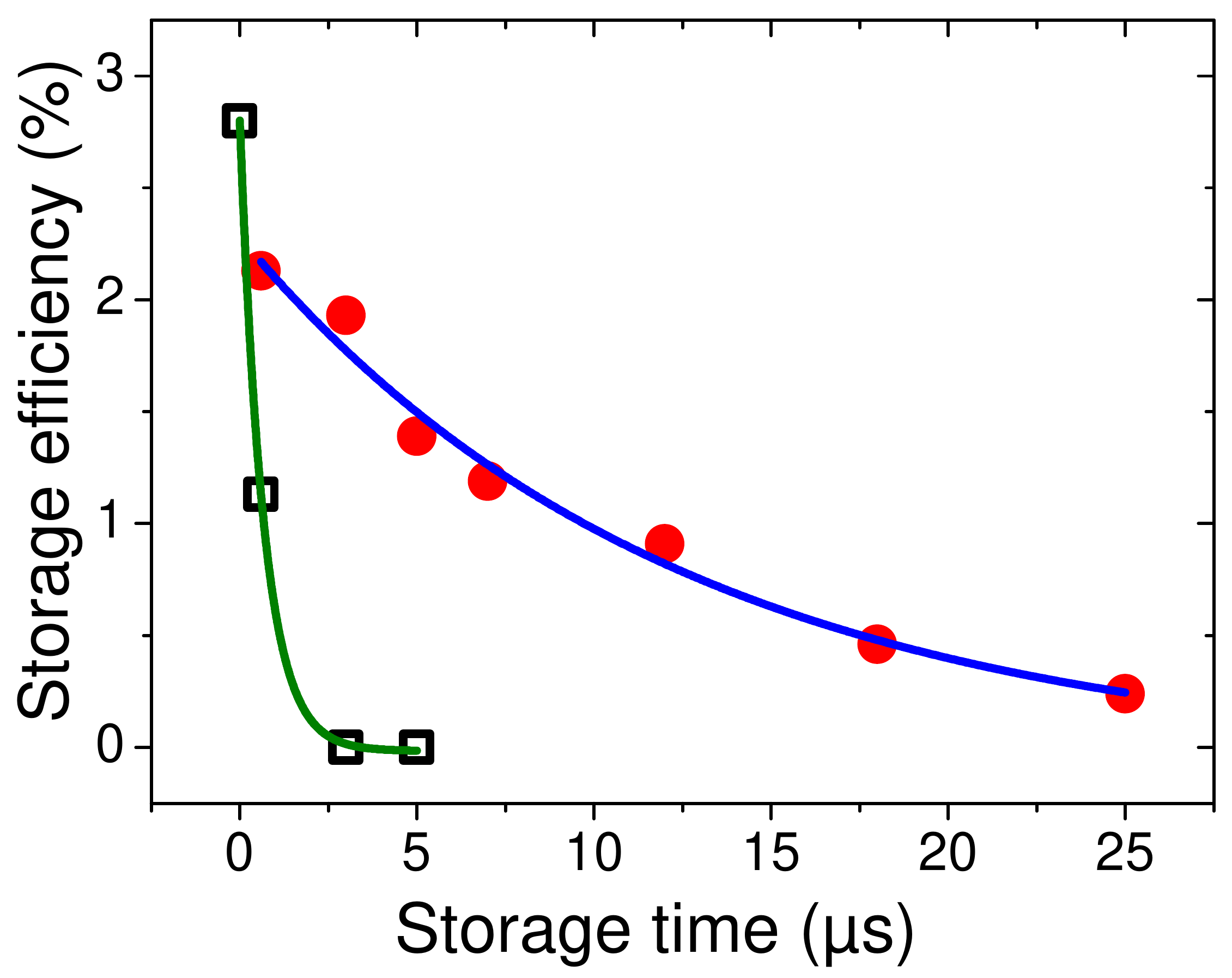} 
\caption{\label{fig5}Measured storage efficiency as a function of the storage time, for a 4\,$\mu$s rise-time exponential pulse when the cell is at the edge of the $\mu$-metal shielding. CPO storage efficiencies (red dots) are recorded in the  lin\,$\perp$\,lin configuration with a B=0.7\,Gauss longitudinal magnetic field. EIT storage (open black squares) measurements are performed in the circ\,$\perp$\,circ configuration with a 1 GHz optical detuning, which gives better efficiencies \cite{Maynard2014}. Continuous lines are exponential fits: when compared to Fig.\,\ref{fig3}, the decay time for CPO storage remains the same (about 10\,$\mu$s) but it strongly decreases down to 0.6\,$\mu$s for EIT storage.}
\end{figure}

We record both the retrieved and the incident probe pulse intensities versus time. The storage efficiency is then obtained by computing the ratio between the areas of these two profiles. Using the small fraction of the coupling beam that leaks through the polarising beamsplitter as a local oscillator, we performed a homodyne detection. One of the mirrors reflecting the coupling beam is mounted on a piezoelectric transducer to modulate the relative phase $\Delta\varphi$ between the local oscillator and the probe pulse (see Fig.\,\ref{fig1}). The detected signal is recorded for many different values of $\Delta\varphi$ and the upper and lower envelopes of the recorded signals correspond to $\Delta\varphi=k.2\pi$ and $\Delta\varphi=\pi+k.2\pi$, where $k$ is an integer. The coupling intensity $I_C$ is measured for each record, and the probe intensity $I_P$ is deduced from the two beam interference formula $I_C+I_P+\alpha\sqrt{I_C I_P}\cos(\Delta\varphi)$. The factor 2 in the interference term is replaced by a factor $\alpha$ to take into account a decrease of contrast, due to a possible small angle between the beams and to their non-planar wavefronts (see reference \cite{Maynard2014}). $\alpha$ is measured for each set of data and found to be larger than 1.7 for the results reported here. The inset in Fig.\,\ref{fig1} shows a typical probe signal after data processing. The first detected peak is the leak transmitted through the cell, due to its finite absorption. After	a 3\,$\mu$s storage time, the coupling beam is switched on again and the retrieved signal is released. Notice that in a real implementation of this protocol for light storage, it would not be necessary to record several sets of data corresponding to several values of $\Delta\varphi$: the CPO memory would work in single shot, just like usual EIT-based memories, if one i) optimizes the quality of the extinction ratio of the polarizers and ii) introduces a small angle between coupling and probe beams in order to be able to detect the probe only.

Storage results are reported in Figs.\,\ref{fig4} and \,\ref{fig5}. All the CPO storage measurements shown here are made at the center of the atomic Doppler profile, with a longitudinal magnetic field, and for a zero probe and coupling beam detuning ($\delta=0$) to select the central CPO resonance [see Figs.\,\ref{fig2}(b) and \ref{fig2}(d)]. Fig.\,\ref{fig4} shows the evolution of CPO and EIT storage efficiencies as a function of the storage time when the cell is at the center of the $\mu$-metal shielding (the magnetic field gradients in the atom cell are negligible). In both cases, one can see an exponential decay with a time constant approximately equal to 10$\mu$s. Since this time constant is much longer than the 98 ns lifetime of the population of the upper level, the measurement of Fig.\ \ref{fig4} shows that the storage with the new protocol discussed here cannot be explained by ordinary CPO involving population oscillations in the upper level. EIT-based storage is known to be limited by the Raman coherence lifetime, but as Fig.\ \ref{fig4} shows the same decay constant for both storage mechanisms, it does not permit to decide whether the new storage investigated here is limited by coherence or population lifetime. In order to lift this indetermination, we purposely degrade the Raman coherence lifetime by adding a magnetic field gradient on the atom cell by pulling it out of its magnetic shielding. The results are shown in fig.\ \ref{fig5}: the EIT storage time drastically decreases, while the storage time of the new mechanism is unaffected. It proves that this last one is not due to a remaining EIT or a coherent Raman process induced by a strong coupling field with a Rabi frequency larger than the Zeeman splitting. Since population oscillations remain unaffected by magnetic field inhomogeneities, we can conclude that the new storage reported here is based on ultra-narrow CPOs \cite{Laupretre2012}. We have also checked that a longitudinal magnetic field does not help to try and decrease random rotations of the spin induced by magnetic field inhomogeneities: indeed, EIT storage in the presence of magnetic field inhomogeneities and a non zero longitudinal magnetic field is very weak and can hardly be detected. Finally, the small differences of CPO storage efficiency levels between Figs.\,\ref{fig4} and \,\ref{fig5} might be explained by small misalignments due to the displacement of the cell and a change in the optical depth (the discharge used to produce metastable helium is slightly modified by the displacement and the field gradients).

\begin{figure}[t]
\center
\includegraphics[width=0.32\textwidth]{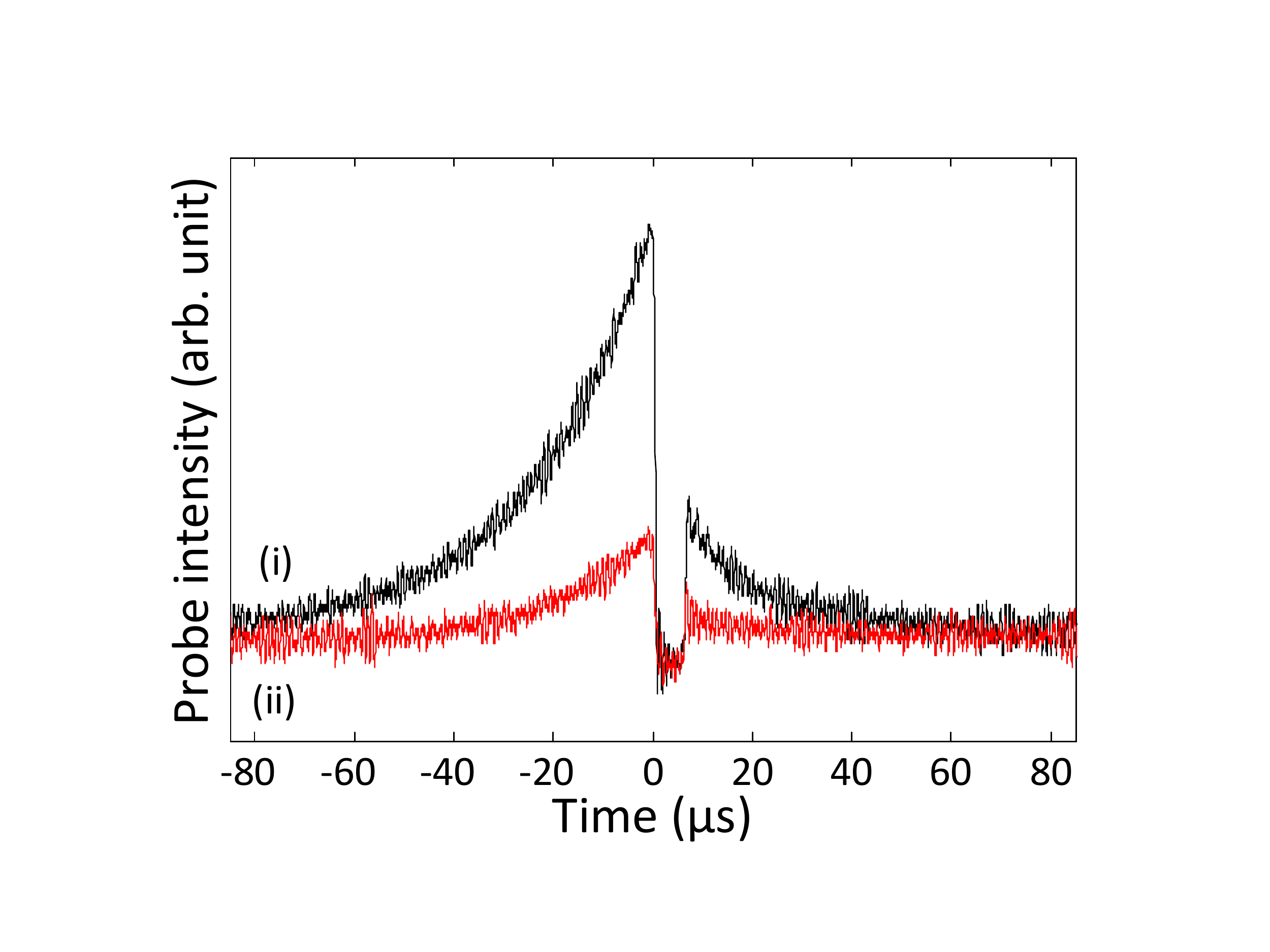} 
\caption{\label{fig6}Interference signal for two different positions of the piezoelectric transducer. The leak and the retrieval interfere in the same way: the black curve shows a constructive interference for both pulses and the grey curve a destructive interference for both pulses.}
\end{figure}

The fact that CPO-based storage preserves the phase is visible using the homodyne detection \cite{Maynard2014}, which measures both the relative phase between the coupling and the leak, and between the coupling and the retrieved pulse. Fig.\,\ref{fig6} shows two signals recorded for (i) constructive interference (black curve) and (ii) destructive interference (red curve) for both leak and retrieved pulses. We checked that when the phase of the coupling beam is scanned, the relative phase is indeed always the same for both the leak and retrieved pulses, which shows that the phase of the probe pulse is preserved during the storage and retrieval process.

In conclusion, we have observed CPO-based storage in a metastable helium gas cell at room temperature, using a $\Lambda$ system selected by polarization. This light storage technique is shown to be phase preserving, and contrary to EIT-based light storage, it is robust to dephasing mechanisms, illustrated here by magnetic field inhomogeneities. The relatively low efficiencies are probably due to the fact that the optical density is lower in the presence of the magnetic field, and the 10\,$\mu$s lifetime of the memory is limited by the transit of the atoms through the laser beam. The efficiency and lifetime of CPO-based storage can thus probably be increased using other motionless $\Lambda$-systems or a broader laser beam. Finally, let us also notice that the idea of substituting a Raman coherence lifetime limited storage timescale by a longer-lived population lifetime limited process was also proposed, using a different approach, in order to increase the efficiency of Raman optical echo based memory \cite{Ham2009}.

The present results open the way to the design of new quantum memories based on solid-state materials that could exhibit $\Lambda$-systems usable at room temperature \cite{Baldit2010}. Theoretical models should be developped understand the limits of such a storage scheme and evaluate the maximum efficiencies that can be obtained. Following the proposal published by \cite{Sharypov2011} about narrowband biphoton sources using CPO in a TLS decaying via a shelving state, CPO in a $\Lambda$-system might also be used for photon pair generation.

\acknowledgments
We thank Jos\'e Tabosa for helpful discussions. The work of M.-A. M. is supported by the D\'elegation G\'en\'erale \`a l'Armement (DGA), France.

\end{document}